\documentclass[conference]{IEEEtran}
\IEEEoverridecommandlockouts
\usepackage{cite}
\usepackage{amsmath,amssymb,amsfonts,bm,multirow,subcaption}
\usepackage{algorithmic}
\usepackage{graphicx}
\usepackage{textcomp}
\usepackage{xcolor}
\usepackage{textgreek}
\usepackage{booktabs}
\usepackage{comment}
\newcommand{\RNum}[1]{\uppercase\expandafter{\romannumeral #1\relax}}
\def\BibTeX{{\rm B\kern-.05em{\sc i\kern-.025em b}\kern-.08em
    T\kern-.1667em\lower.7ex\hbox{E}\kern-.125emX}}
\begin{document}

\title{Mixed-Variable PSO with Fairness on Multi-Objective Field Data Replication in Wireless Networks\\

}

\author{\IEEEauthorblockN{Dun Yuan$^{1+}$}
\and
\IEEEauthorblockN{Yujin Nam$^{2+}$}
\and
\IEEEauthorblockN{Amal Feriani$^1$}
\and
\IEEEauthorblockN{Abhisek Konar$^1$}
\and
\IEEEauthorblockN{Di Wu$^{1*}$}
\and
\IEEEauthorblockN{Seowoo Jang$^2$}
\and
\IEEEauthorblockN{Xue Liu$^1$}
\and
\IEEEauthorblockN{Greg Dudek$^1$}
\and

\centerline{$^1$Samsung AI Center, Montreal, QC H3A 3G4, Canada} \\
\centerline{$^2$Samsung Electronics, Korea (South)}\\
\centerline{(e-mails: firstname.lastname@samsung.com, dun.yuan@partner.samsung.com, }\\
\centerline{di.wu1@samsung.com, yujin\_.nam@samsung.com)}\\
\centerline{$^+$ Equal Contribution. $^*$Corresponding author}}

\maketitle

\begin{abstract}
Digital twins have shown a great potential in supporting the development of wireless networks. They are virtual representations of 5G/6G systems enabling the design of machine learning and optimization-based techniques. Field data replication is one of the critical aspects of building a simulation-based twin, where the objective is to calibrate the simulation to match field performance measurements. Since wireless networks involve a variety of key performance indicators (KPIs), the replication process becomes a multi-objective optimization problem in which the purpose is to minimize the error between the simulated and field data KPIs. Unlike previous works, we focus on designing a data-driven search method to calibrate the simulator and achieve accurate and reliable reproduction of field performance. This work proposes a search-based algorithm based on mixed-variable particle swarm optimization (PSO) to find the optimal simulation parameters. Furthermore, we extend this solution to account for potential conflicts between the KPIs using \textalpha-fairness concept to adjust the importance attributed to each KPI during the search. Experiments on field data showcase the effectiveness of our approach to (i) improve the accuracy of the replication, (ii) enhance the fairness between the different KPIs, and (iii) guarantee faster convergence compared to other methods. 


\end{abstract}


\section{Introduction}

Future wireless generations (i.e., beyond 5G) promise to provide faster connectivity to support novel data-intensive applications like extended reality and metaverse. Artificial intelligence (AI) serves as one of the core technologies for these future networks. In this context, reliable virtual representations of the wireless system or digital twins (DTs) are crucial to designing and assessing AI-based solutions for future field deployment~\cite{khan_digital-twin-enabled_2022}. DTs can emulate the physical wireless system and estimate its future behavior for other downstream tasks, such as reinforcement learning (RL)\cite{9855432}. They also support the industrial management and automation of wireless network systems\cite{9838860, 9838407}.

As the name ``twin" suggests, DTs need to be as consistent with their physical counterparts as possible in terms of observed data and performance measurements\cite{jiang_industrial_2021}. Achieving this consistency is the primary purpose of the data replication problem. For the rest of the paper, we focus on a simulation-based twin where a simulator replicates all the low-level components of a radio access network. The data replication problem involves calibrating the simulator input parameters (e.g., traffic parameters, user-specific parameters, modulation, coding scheme) to minimize the discrepancies between the simulated key performance indicators (KPIs) and the field ones. This problem is challenging for two main reasons: (i) the parameter search space is in high dimension and potentially with mixed variable types (i.e., discrete and continuous), and (ii) with multiple KPIs of interest, the replication task takes the form of a multi-objective problem with potential conflicts amongst the KPIs. This work mainly focuses on the second challenge, where we propose a solution to ensure a fair replication of the different KPIs. A fair replication means that replication errors for all KPIs are reduced with similar consideration. We base our solution on a recent particle swarm optimization (PSO) algorithm, coined mixed-variable PSO \cite{wang_particle_2021}, that handles mixed search space, unlike the standard PSO algorithm \cite{kennedy_particle_1995}.   Machine learning has been widely used for different types of real-world application~\cite{zhang2022metaems,wu2018machine,uang2021modellight,fucloser,fu2022reinforcement,wu2019multiple,DBLP:conf/ijcai/LinW21,wu2022efficient,wu2022short}. It has also been application to different types of communication use cases~\cite{feriani2022multiobjective,ma2022coordinated,li2022traffic}. With the proposed method in this work, we can better support the training of machine learning models.

Multi-objective problems are typically solved using scalarization. Scalarization is a technique where the different objectives are weighted according to a predefined preference. The preference over the objectives may not be known prior to the learning process. We argue that hand-picking the preference is challenging and can hinder the final replication performance. A KPI with a higher weight will skew the replication process by introducing heavy bias that favors that particular KPI over the others—like in the network resource allocation problem, maintaining fairness across the system is critical\cite{ferragut2013network}. Ignoring KPIs with limited preference might harm the field application of DTs. To overcome this issue, we use \textalpha-fairness, a generalized framework that controls the trade-off between the replication error and fairness \cite{pmlr-v80-komiyama18a} (i.e., proportional fairness and max-min fairness depending on the value of \textalpha). Therefore, given a preference, the data replication problem involves optimizing the \textalpha-fairness objective function.

To assess the effectiveness of our approach, we conducted extensive experiments using field data from different geographic areas. We evaluated our solution using metrics such as replication error, fairness, and convergence rate. The proposed framework outperforms different baselines regarding replication accuracy and convergence rate while guaranteeing better fairness between objectives. 
To summarize, this paper has the following contributions:

\begin{itemize}
  \item We model the data replication for DTs as a multi-objective optimization problem;
  \item We propose a novel and general solution based on mixed-variable PSO to handle hybrid search spaces and \textalpha-fairness to balance the performance between KPIs;
  \item We validate our approach on \emph{field} performance data and showcase that the proposed framework improves the replication performance, the fairness between KPIs, and the convergence rate.
\end{itemize}

The remainder of this paper is organized as follows. Section \RNum{2} gives a brief survey about works in related areas. Section \RNum{3} formalizes the problem in terms of mathematical representations. Section \RNum{4} introduces the methods proposed in this paper. Section \RNum{5} describes the setup for the experiments and baseline methods. Section \RNum{6} concludes the idea and results of the paper.

\section{Related Work}

\textbf{Digital Twins}. The original idea of DTs was introduced decades ago\cite{grieves_digital_2014}. Researchers described DTs as a virtual representation of an actual physical instance\cite{rosen_about_2015}. Although the concept has existed for a long time, the trend of applying DTs in industries has proliferated only in recent years\cite{jones_characterising_2020}. DTs' applications often require large-scale autonomous systems to keep them practical and productive\cite{rosen_about_2015}. It is introduced in \cite{khan_digital-twin-enabled_2022} how DTs enable the research and deployment of 6G networks. The authors also reviewed DTs' taxonomy, challenges, and opportunities in wireless network systems. The work offered a comprehensive survey and overview of DTs' key concepts and applications. Notably, they highlighted the importance of reliable field data replication for adopting DTs in the industrial domain.

\textbf{Particle Swarm Optimization}. 
The PSO algorithm was introduced by \cite{kennedy_particle_1995} in 1995, and the standard format is simple in its velocity update functions. It is an effective method used in many applications. Further improvements to the original algorithm and its applications are proposed in \cite{kennedy_particle_1997,shi_modified_1998} by the original authors.
Subsequently, a different line of work suggested improvements and extensions to tackle different shortcomings of the standard PSO algorithm. In particular, a mixed-variable variant of PSO was recently proposed \cite{wang_particle_2021} to handle both discrete and continuous variables in the search space since the original PSO only works for continuous search spaces. We refer the reader to \cite{poli_analysis_2008} for an extensive survey of the different PSO variants and their area of applications.


\textbf{Fairness}. Fairness in network engineering ensures equitable distribution of resources to the connected users \cite{douligeris_fairness_1995}. In research problems such as network resource allocation, fairness is an essential metric that needs consideration. Reference \cite{shi_fairness_2014} presented an overview of the use of fairness in wireless networks. There are two main approaches to incorporating fairness. The first one relies on fairness measures that map resource allocation vectors to a real number, such as Jain's index~\cite{jain_quantitative_1984} or an entropy function. The second approach consists of constructing a fairness function that can be integrated into the optimization objective. \textalpha-fairness, for instance, is one of the more popular fairness objective functions. It was introduced to the network research community by \cite{kelly_rate_1998, mo_fair_2000}. Reference \cite{lan_axiomatic_2010} presented five axioms of fairness measures to unify these approaches for incorporating fairness. It also explained the properties and usage of \textalpha-fairness, especially the effect of chosen $\alpha$ value on the trade-off between efficiency and fairness.



\section{Problem Formulation}


We consider a simulation-based twin that aims to replicate 5G wireless network system state and behavior. The simulator emulates a homogeneous multi-band downlink Orthogonal Frequency Division Multiple Access (OFDMA) cellular network consisting of $B$ 3-sector macro-gNodeBs (gNBs) spaced by an inter-site distance $d$. Each sector of the gNB operates over $K$ frequency bands. A user is served by a single frequency band $k$ from a given sector $j$ of a particular macro-gNB $b$. We assume that the frequency bands and their associated bandwidths are homogeneous for the whole network. The virtual gNBs are configured to match their real counterparts regarding transmission powers, antenna gains, bandwidths, etc. Other parameters can be set according to the available field data, such as the channel parameters (the 256 QAM usage ratio and the average number of codewords). In addition to the aforementioned parameters, other simulation parameters must be determined to reduce the gap between the simulated and field KPIs. 
\begin{figure}[htbp]
\centerline{\includegraphics[scale=0.34]{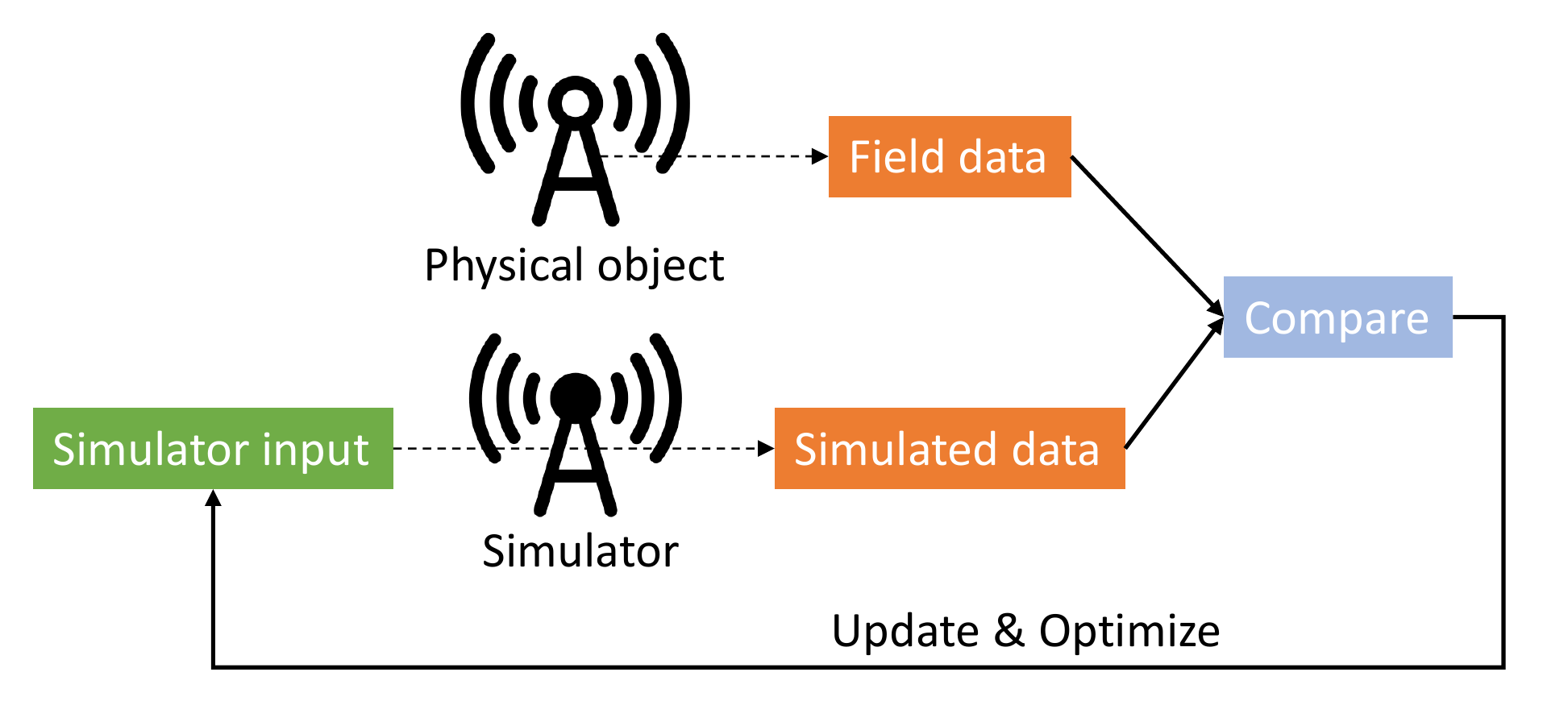}}
\caption{The pipeline of field data replication for a simulation-based twin}
\label{pipeline}
\end{figure}

Note that this paper assumes the simulator to be a black box system in terms of its inputs and outputs. No prior knowledge of the dynamics in the field data is assumed within the design of the data replication procedure illustrated in Fig.~\ref{pipeline}. 
Formally, the data replication problem is given by:
\begin{equation}
\begin{aligned}
\min_{\bm{x}} \quad &  
\bm{G} = \left[g(f(\bm{x})_1,y_1),\dots, g(f(\bm{x})_n,y_n)\right] \\
{s.t.} \quad & \bm{x}\in\bm{S}.
\end{aligned}
\end{equation}

where $\bm{x}$ represents the vector of simulation parameters, $\bm{y}$ represents the vector of field KPIs, $f$ represents the simulator as a black box function, $f(x)$ is the vector of simulated KPIs, $g$ represents the evaluation function, $\bm{S}$ represents the search space of $\bm{x}$ and $n$ is the number of KPIs.




This paper uses the scalarization method to simplify the problem into a single-objective one. One way to consider the weights for scalarization is simply the no-preference method, which assumes no preference exists for different KPIs (i.e., KPIs are weighted equally). However, provided by the decision makers in actual scenarios, preference exists for some KPIs over others. The decision makers might fail to provide particular and realistic normalized weights over the different KPIs but an expression of the preferences. Assuming that the preferences could be formulated as a vector $\bm{p}=[p_1,\dots,p_n]$, the scalarized approach consists in multiplying the objectives with $\bm{p}$ as follows:
\begin{align}\label{eq:pb}
\min_{\bm{x}} \quad &  
\bm{p}\cdot \bm{G} = \sum_{i=1}^n p_i\cdot g(f(\bm{x})_i,y_i)
\quad \textrm{s.t.} \quad \bm{x}\in\bm{S}.
\end{align}

To solve the problem in (\ref{eq:pb}), we start by presenting the adopted search algorithm. Afterward, we will detail the \textalpha-fairness solution to achieve a good trade-off between the replication performance and the fairness between KPIs.


\section{Proposed Method}
\subsection{Search Algorithms}
\subsubsection{Standard PSO}\label{sec:pso}
The standard PSO algorithm \cite{kennedy_particle_1995} uses $X_i=(x_i^1,x_i^2,...,x_i^D)$ to denote the position of the $i$th particle in a $D$-dimensional search space. The velocity of particle $i$ is written as $V_i=(v_i^1,v_i^2,...,v_i^D)$. The velocity update function of the standard PSO algorithm is then formulated as follows:
\begin{equation}
\begin{aligned}
V_i(t+1)&=w\cdot V_i(t)+c_1r_1(pb_i(t)-X_i(t)) \\
&\quad +c_2r_2(gb(t)-X_i(t))
\end{aligned}
\end{equation}
where $pb_i$ represents the personal best position of particle $i$, and $gb$ represents the global best position that has been explored by all of the particles. Through the configuration of $c_1$ and $c_2$, the PSO algorithm could be modified to achieve the trade-off between cognitive behaviors and social behaviors. After the velocity update, the position vector of the particles is updated as follows: 

\begin{equation}
    X_i(t+1)=X_i(t)+V_i(t).
\end{equation}

Although the standard PSO algorithm has been applied in many fields, it still suffers several drawbacks. It can be used only for continuous search spaces and does not show outstanding performance in exploitation. Research endeavors extended the algorithm to work for binary selection problems \cite{kennedy_particle_1997}, but mixed-variable search spaces remain a tricky problem for PSO-based algorithms. In this paper, we do not restrict the search space to be continuous and consider both continuous and discrete simulation parameters. We utilize a recent extension of the PSO algorithm called mixed-variable PSO.  

\subsubsection{Mixed-variable PSO}

Mixed-variable PSO introduces a new encoding scheme and reproduction method to handle both continuous and discrete variables\cite{wang_particle_2021}. To do so, they assume that the search space contains $Z$ continuous parameters and $L$ discrete ones,  and the position of a particle $i$ can be encoded as in $X_i=(x_i^1,x_i^2,...,x_i^Z,x_i^{Z+1},x_i^{Z+2},...,x_i^{Z+L})$.
This representation means that different input parameters of the particle $i$ are still integrated into only one position vector $X_i$. However, the continuous parameters and discrete parameters are organized in different parts. Therefore, mixed-variable PSO can handle the update process through this encoding scheme without introducing additional vectors.
\vspace{-10px}
\begin{figure}[htbp]
\centering
\includegraphics[scale=0.52]{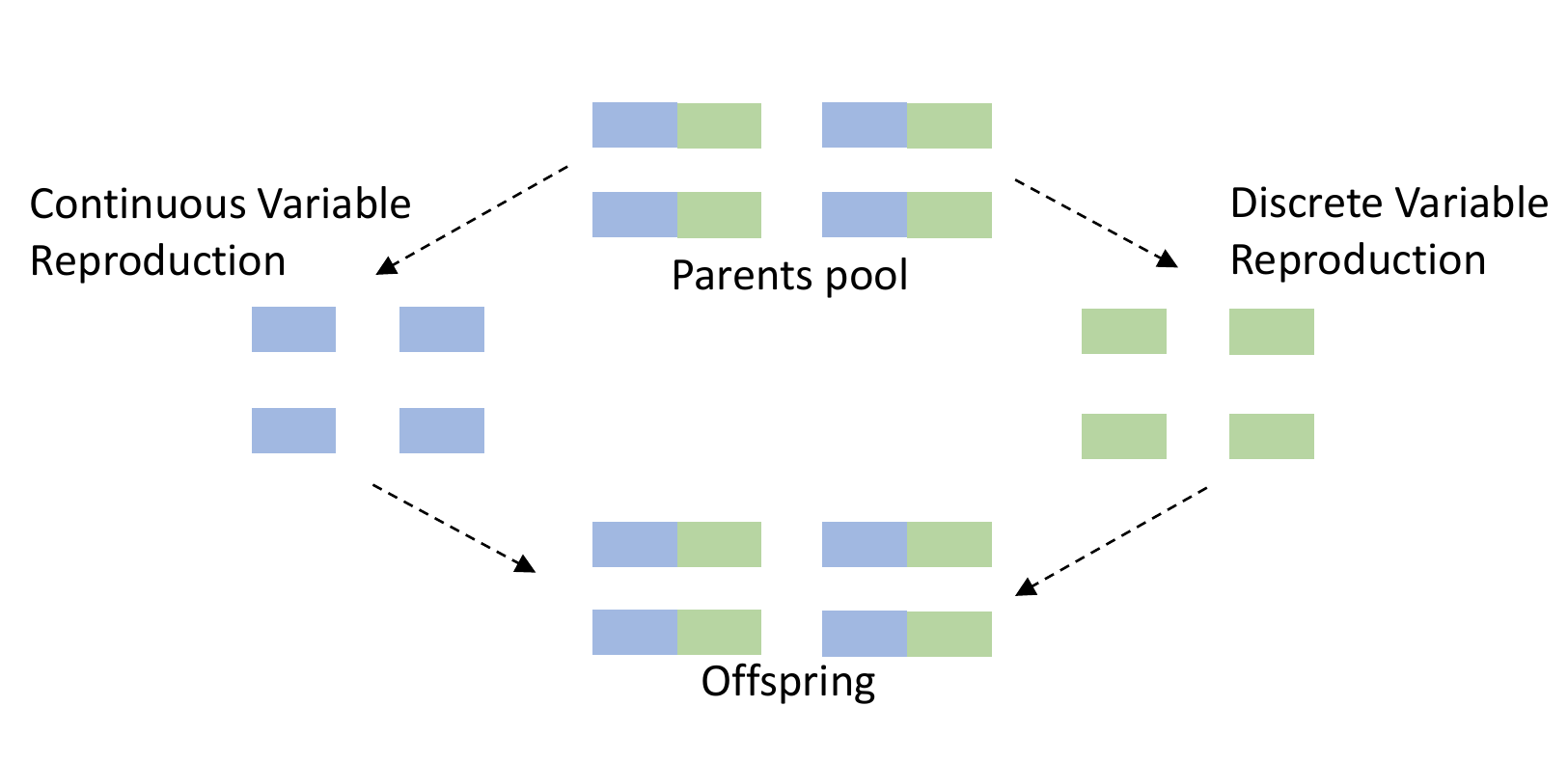}
\caption{Illustration of the reproduction method in mixed-variable PSO algorithm}
\label{reproduction}
\end{figure}
Fig.~\ref{reproduction} illustrates the reproduction method based on the mixed-variable encoding scheme previously introduced. First, the parents pool includes the position of all particles, which consists of two parts - the blue part represents the continuous parameters $x_i^1,x_i^2,...,x_i^Z$ and the green part represents the discrete parameters $x_i^{Z+1},x_i^{Z+2},...,x_i^{Z+L}$. Since both continuous-specific and discrete-specific reproduction methods are implemented with different implementations, the reproduction processes are conducted separately. The reproduced segments are combined into new complete offspring, which stands for the updated positions of the particles in the mixed-variable PSO algorithm. The reproduction processes can also be done simultaneously, enabling better efficiency in the optimization process.

The mixed-variable PSO algorithm enables the ability to find the simulation parameters through several update iterations in our dataset that has both continuous and discrete variables. The following section explains how we augmented this search algorithm to account for the trade-off among the different KPIs.

\subsection{Data replication with Fairness}

Fairness in the network research area is usually conducted in two different ways — the first consists of evaluating the data or the solution to a problem with available fairness metrics. The second technique uses a composition function that defines a new objective function for the optimization problem. In this paper, we adopt the second approach.

To mitigate the unfairness emerging from optimizing different KPIs, this paper proposes a simple and easily deployed idea based on \textalpha-fairness. To do so, we extend the standard data replication problem in (\ref{eq:pb}) to \emph{jointly} optimize for the replication performance and fairness. We use $T_i=p_i\cdot g(f(\bm{x})_i,y_i)$ to denote the objective value in (\ref{eq:pb}).
\begin{equation}
\begin{aligned}
    \min_{\bm{x}} \quad & \sum_{i=1}^n U_\alpha(T_i) \\
    & \text{where } U_\alpha(T_i) = \begin{cases}
      \frac{T_i^{1-\alpha}}{1-\alpha}, & \alpha\geq0, \alpha\neq1 \\
      \log{T_i}, & \alpha=1
    \end{cases}.
\end{aligned}
\end{equation}
    


This new problem formulation transforms the objective function depending on the chosen $\alpha$ value within $[0,\infty)$. 
Choosing the correct value of $\alpha$ is important when applying \textalpha-fairness, as a larger $\alpha$  prioritizes fairness over efficiency \cite{lan_axiomatic_2010}.
In this paper, we choose $\alpha=1$ to achieve proportional fairness, meaning that increasing a certain proportion of one value cannot be based on decreasing a more significant proportion of other values. After applying the concept of \textalpha-fairness to the scalarized multi-objective problem in (\ref{eq:pb}), the overall fairness of objectives on different KPIs is improved.


\section{Experiments}

\subsection{Setup}

In this paper, we consider $5$ sites located in different geographic areas. A full buffer traffic model is assumed, as in 3GPP TR.36814. The packet size and the mean of the inter-file arrival distribution at time $t$ is denoted by $s_t$ and $\mu_t$ respectively. 
At the beginning of the simulation, the user equipment (UEs) are uniformly distributed geographically across the gNBs. During the simulation, they either remain static or undergo a random motion with a constant velocity. We also consider three different UE states: idle, inactive, and active. Only active UEs receive data transmissions from the gNBs. Without loss of generality, we evaluate our method on the central sector of the gNB. We optimize for two frequency bands $f_1$ and $f_2$ where $f_1 < f_2$ and report the results for these specific bands. Table \ref{tab:sim-params} lists the set of simulation parameters that remain constant across all the experiments.
\begin{table}[h!]
\centering
\caption{\label{tab:sim-params} Simulation parameters}
\begin{tabular}{*2l}
\toprule
Parameters &  Values\\
\midrule
Path-loss & $128.1 + 37.6 \log_{10}(d)$ \\
Traffic model & Full Buffer \\
Scheduler & Proportional Fair \\ 
UE velocity & $3$ m/s \\
UE mobility model & Static/random motion \\
Thermal noise & $-174$ dBm/Hz\\
\bottomrule
\end{tabular}
\end{table}

\subsection{Dataset}
The field datasets contain KPI measurements collected over a span of 24 hours broken into intervals $\delta_t = 15$ mins. We consider the following KPIs for each carrier frequency:
\begin{itemize}
    \item \textbf{Active UEs}: is the average number of active UEs over the period $\delta_t$;
    \item \textbf{Cell load}: is the physical resource block utilization ratio of a given band;
    \item \textbf{Downlink volume}: is the average amount of successfully transmitted data packets in the downlink.   
\end{itemize}

\subsection{Search Space}
Our search space includes the traffic parameters $s_t$, $\mu_t$ and the number of UEs per cell (i.e., a cell refers to a sector and frequency pair). Table~\ref{tab:search-params} summarizes the parameters, their type, and the search boundaries. Note that we consider both continuous and discrete parameters.   

\begin{table}[h!]
\centering
\caption{\label{tab:search-params} Search Space}
\begin{tabular}{lll}
\toprule
Parameter & Type & Range \\
\midrule
Packet size $s_t$ & Continuous & $[0.05,30]$ kbytes\\
Inter-file arrival mean $\mu_t$ & Continuous & $[0,300]$ ms  \\
Number of UEs per cell & Discrete & $[3,50]$ \\
\bottomrule
\end{tabular}
\end{table}



\subsection{Baselines}

In this work, we consider three search-based baselines. All methods are repeated for 50 iterations. The PSO-based methods use 5 particles as the swarm, and $w=1.1$, $c_1=1.1$, $c_2=0.8$ as values of hyperparameters.

\begin{itemize}
    
    \item \textbf{Random search} \cite{DBLP:journals/jmlr/BergstraB12}: is a common approach in which the parameters are selected randomly and independently. It is simple and easy to implement. However, it can be time-consuming;
    \item \textbf{Bayesian Optimization} (BO) \cite{mockus_bayesian_2012, eggensperger2013towards}: is an iterative approach where the next parameters are determined based on the previous evaluations. BO algorithms consist of two main components: (i) a surrogate function that models the posterior distribution of the objective function (e.g., $g$) using the observed data points, and (ii) an acquisition function that suggests the next parameters evaluated based on the surrogate function while maintaining a trade-off between exploitation and exploration;
    \item \textbf{Standard PSO} \cite{kennedy_particle_1995, kennedy_particle_1997}: is the traditional PSO method as explained in Section \ref{sec:pso}. We use a round-off method for discrete parameters.
\end{itemize}
\begin{figure*}[ht]
\centering
    \begin{subfigure}[t]{0.3\textwidth}
        \includegraphics[width=\textwidth]{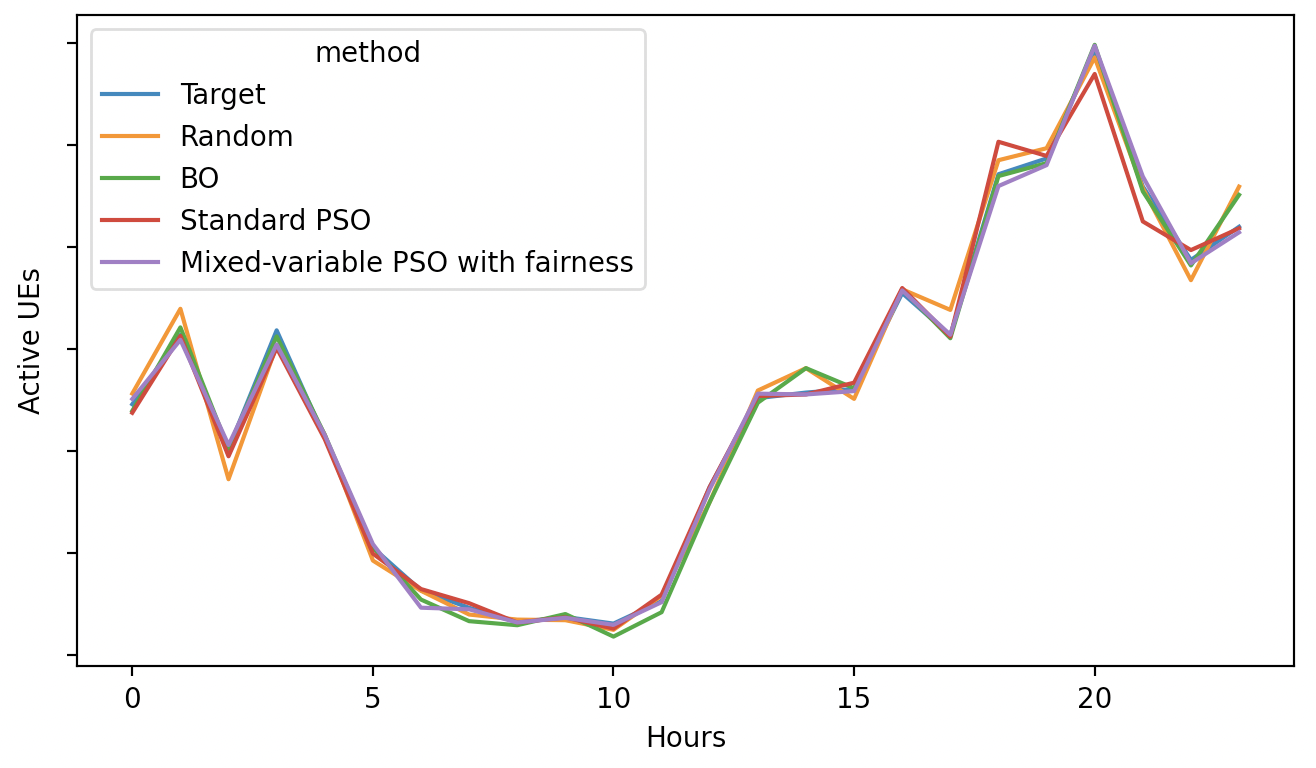}
        \caption{Active UEs (0.8)}
        \label{KPIs_1}
    \end{subfigure}
    \begin{subfigure}[t]{0.3\textwidth}
        \includegraphics[width=\textwidth]{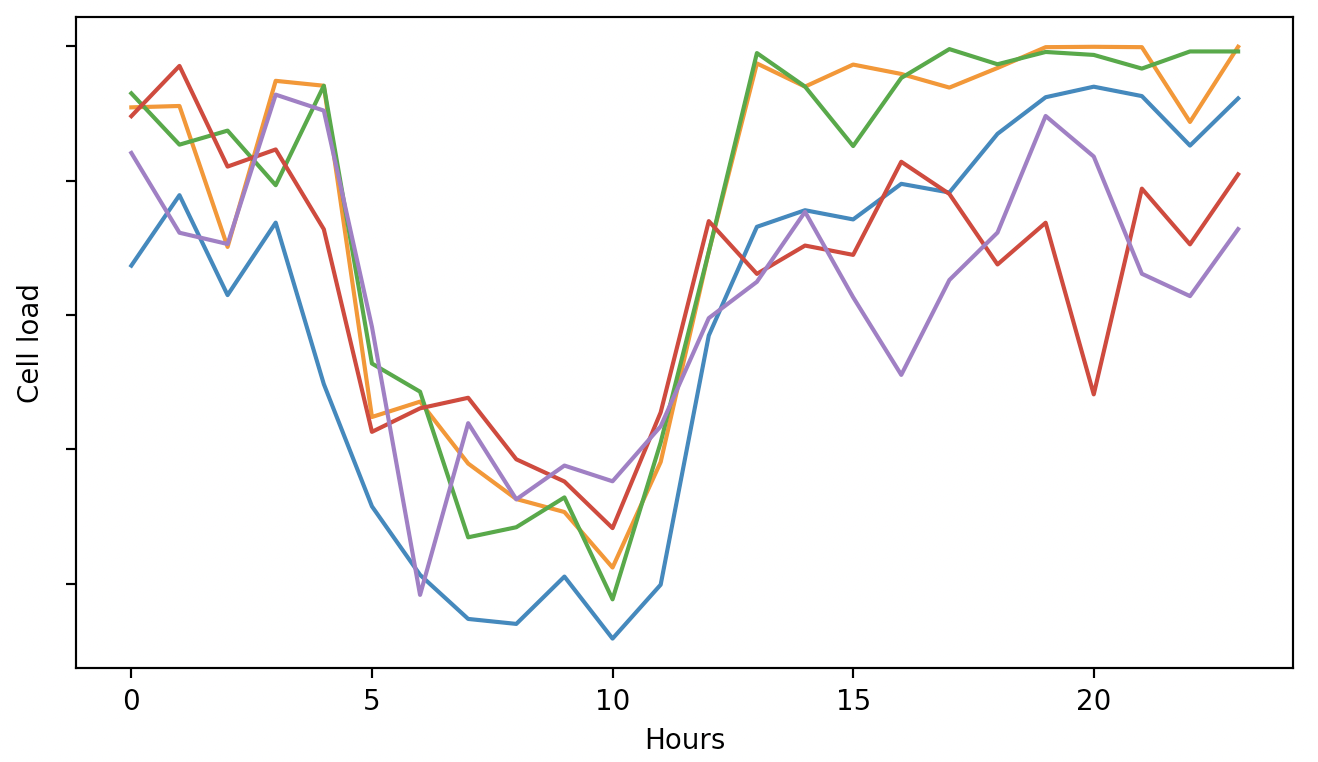}
        \caption{Cell load (0.1)}
        \label{KPIs_2}
    \end{subfigure}
    \begin{subfigure}[t]{0.3\textwidth}
        \includegraphics[width=\textwidth]{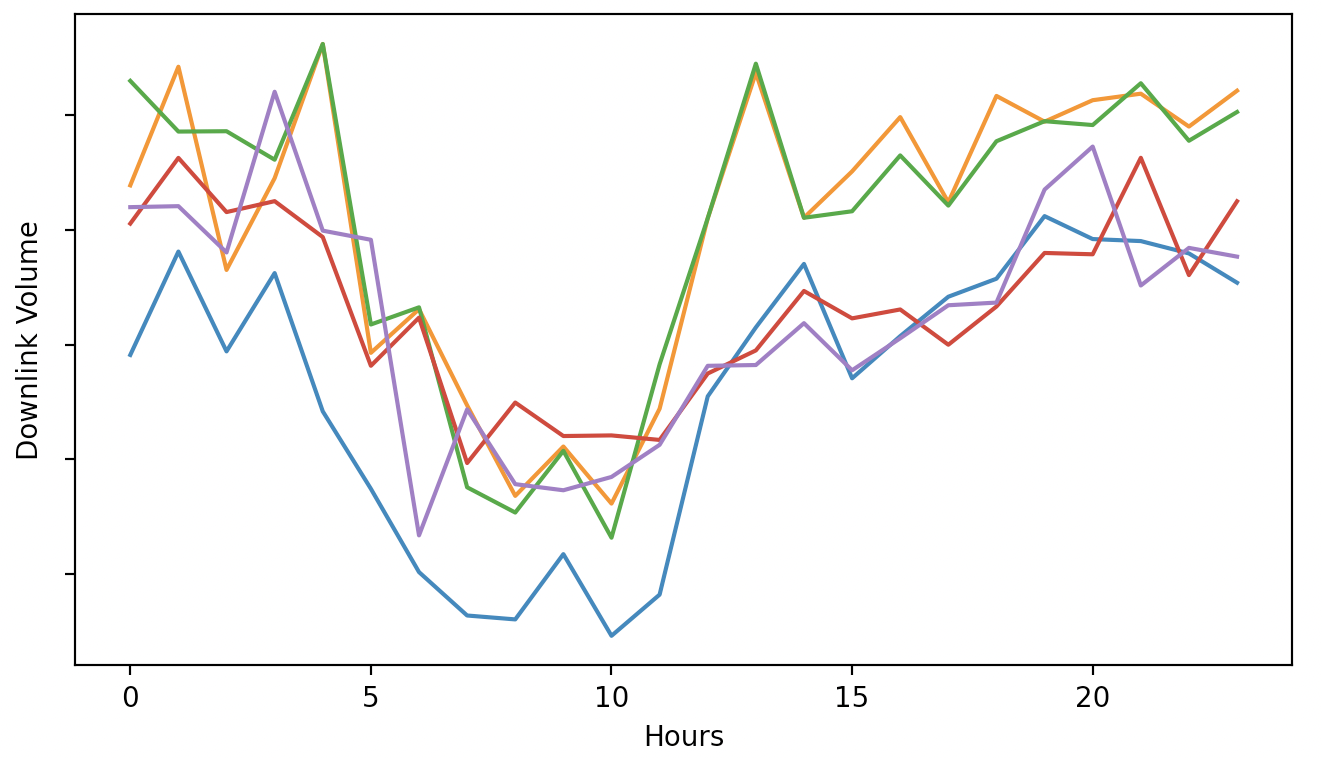}
        \caption{Downlink Volume (0.1)}
        \label{KPIs_3}
    \end{subfigure}
\caption[Caption for LOF]{An illustrative example of the simulated and field KPIs using $\bm{p} = [0.8,0.1,0.1]$. A higher weight on the active UE KPI led to a better replication at the expense of the other KPIs.\protect\footnotemark}
\label{Comparison with target data}
\end{figure*}



\subsection{Evaluation Procedure}
To investigate the impact of the choice of the preferences on the replication outcome, we consider three different weight vectors: $\bm{p}_1 = [0.8,0.1,0.1]$, $\bm{p}_2 = [0.1,0.8,0.1]$ and $\bm{p}_3 = [0.1,0.1,0.8]$. Note that each weight vector places more emphasis on one of the KPIs. We expect that the performance of the baselines will be biased towards the KPI with more weight, whereas our method will reach a better trade-off between all KPIs. For a more comprehensive comparison, we consider three evaluation metrics to assess different aspects of the proposed method and the baselines: 
\begin{itemize}
    \item \textbf{Replication accuracy}: to assess the quality of the replication, we use the mean absolute percentage error (MAPE) as the metric to calculate the closeness of simulated KPIs to field target ones.
    \begin{equation}
        \text{MAPE}=\frac{100\%}{n}\sum_{i=1}^{n}{|\frac{f(x)_i-y_i}{y_i}|}.
    \end{equation}
        
    \item \textbf{Jain's index} \cite{jain_quantitative_1984}: to measure the fairness between the different KPIs and show the advantage of taking fairness into consideration during the optimization process.
    \begin{equation}
        \mathcal{J}(\bm{T}=\bm{p}\cdot \bm{G})=\frac{(\sum_{i=1}^n{T_i})^2}{n\cdot\sum_{i=1}^n{{T_i}^2}}.
    \end{equation}

    \item \textbf{convergence rate}: to quantify the efficiency of the competing methods. It is the number of iterations needed for the objective value to be close to the optimal value.
\end{itemize}



\section{Results}

\subsection{Replication accuracy}
\footnotetext{The data presented is a variation of the field data. Y-axis labels are removed due to confidentiality requirements.}
We first examine the average accuracy across the different sites, frequencies, and preferences. From Table~\ref{Average MAPE of different methods}, we observe that our proposed method outperforms the baselines. The random search method has the worst performance in terms of MAPE, especially its standard deviation. BO yields a better replication than the standard PSO algorithm.
\begin{table}[htbp]
\caption{Average MAPE of different methods}
\begin{center}
\begin{tabular}{l c}
\toprule
\textbf{Method}&\textbf{MAPE}\\
\midrule
Random                              &35.34\textpm33.44\\
\midrule
BO                                  &20.76\textpm8.36\\
\midrule
Standard PSO                        &22.54\textpm5.43\\
\midrule
Mixed-variable PSO with fairness    &\textbf{20.39\textpm6.92}\\
\bottomrule
\end{tabular}
\label{Average MAPE of different methods}
\end{center}
\end{table}

Table~\ref{MAPE of different KPIs} breaks down the detailed MAPE results of each KPI for three different preference vectors. We observe that the performance of all baselines is biased towards the KPI with the highest importance weight. However, 
mixed-variable PSO with fairness achieves a trade-off between the MAPE of the preferred one and the other two KPIs, which shows that our proposed method focuses more on optimizing the KPIs that are not preferred. Fig.~\ref{Comparison with target data} shows the comparison among all the methods and the field target data using one of the experiment settings. Note that preference vector $\bm{p}$ was set to $[0.8,0.1,0.1]$, lending a higher emphasis on optimizing for Active UEs.
 
\begin{table}[htbp]
\caption{MAPE of different KPIs}
\begin{center}
\resizebox{\columnwidth}{!}{
\begin{tabular}{l|ccc}
\hline
&\multicolumn{3}{c}{\textbf{MAPE}}\\
\cline{2-4}
\textbf{Method}& Active UEs& Cell load & Downlink Volume\\
\cline{2-4}
&\multicolumn{3}{c}{$\bm{p}_1=[0.8,0.1,0.1]$}\\
\hline
Random & 12.25&64.25&88.02\\
BO &\textbf{7.08}&62.30&84.75\\
Standard PSO & 11.59&66.30&83.35\\
Mixed-variable PSO with fairness & 10.29 & \textbf{45.76}&\textbf{68.92}\\
\hline
&\multicolumn{3}{c}{$\bm{p}_2=[0.1,0.8,0.1]$}\\
\hline
Random&54.02&15.68&73.73\\
BO &44.66&15.72&74.62\\
Standard PSO &53.37&14.98&82.18\\
Mixed-variable PSO with fairness &\textbf{40.93}&\textbf{13.67}&\textbf{71.67}\\
\hline 
&\multicolumn{3}{c}{$\bm{p}_3=[0.1,0.1,0.8]$}\\
\hline
Random &62.24&61.00&10.41\\
BO &54.10&53.14&\textbf{4.95}\\
Standard PSO &59.26&49.02&6.73\\
Mixed-variable PSO with fairness &\textbf{44.85}&\textbf{42.29}&8.04\\
\hline
\end{tabular}}
\label{MAPE of different KPIs}
\end{center}
\end{table}

\subsection{Fairness metric}

Table~\ref{Results with Jain's index} shows the fairness achieved by the different methods measured by Jain's fairness index. From the results, we conclude that our proposed method shows a better capability of balancing the trade-off among the different KPIs leading to solutions with greater fairness.
\begin{table}[!h]
\caption{Fairness results of different methods with Jain's index}
\begin{center}
\begin{tabular}{lc}
\toprule
\textbf{Method}&\textbf{Jain's index}\\
\midrule
Random                              &0.77\textpm0.11\\
\midrule
BO                                  &0.74\textpm0.05\\
\midrule
Standard PSO                        &0.75\textpm0.04\\
\midrule
Mixed-variable PSO with fairness    &\textbf{0.80}\textpm0.02\\
\bottomrule
\end{tabular}
\label{Results with Jain's index}
\end{center}
\end{table}
\subsection{convergence rate}
\begin{figure}[!htbp]
\centerline{\includegraphics[scale=0.4]{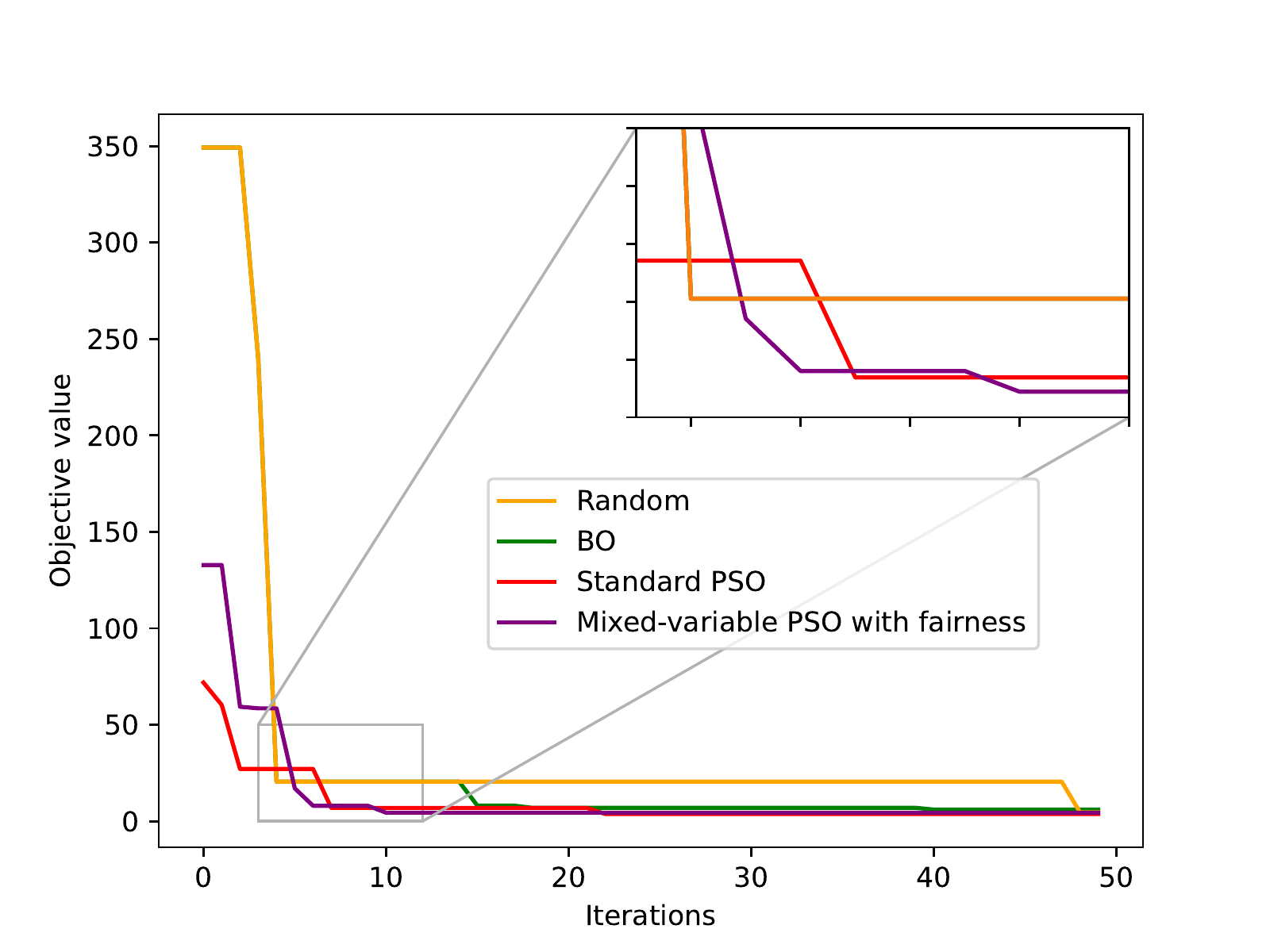}}
\caption{Comparison of objective values of different methods over 50 iterations in a specific setting. BO and random search have identical objective values for 14 iterations.}
\label{convergence rate comparison}
\end{figure}
The convergence rate of the different methods is shown in Fig.~\ref{convergence rate comparison}. Standard PSO starts with a lower objective value and converges in the 8th iteration. Mixed-variable PSO with fairness has an even faster convergence rate -- it converges in the 7th iteration. One reason PSO-based methods work well in terms of convergence rate is that multiple particles in PSO-based search significantly help the exploitation aspect of the search, enabling it to find the optimal value within fewer iterations.

\section{Conclusion}
Field data replication for simulation-based twins is a multi-objective problem with different KPIs. 
The main contributions of this paper are in three aspects: (i) formulating the problem mathematically to clarify, (ii) using \textalpha-fairness to address this multi-objective optimization problem for simplification, and (iii) applying a mixed-variable version of PSO to increase the performance. This paper validates the proposed approach by conducting experiments on field measurement datasets. According to the results, our proposed method outperforms the three baseline methods in terms of MAPE with field target data, fairness metrics, and convergence rate. In this paper, the formulation and experiments focused on different KPIs averaged over all UEs, which can be extended to the UE level in future works. More search methods can also be applied to increase the capability of exploration during the optimization process. 



\bibliographystyle{IEEEtran}
\bibliography{IEEEabrv,refs}

\end{document}